# Sensitivity enhancement in optical micro-tube resonator sensors via mode coupling

Tao Ling[a], and L. Jay Guo[b]

*Department of Electrical Engineering and Computer Science, The University of Michigan, Ann Arbor, Michigan 48109, USA*

A liquid filled, low refractive index material inner-coated silica micro-tube has been proposed and studied as a coupled micro-resonator sensor to greatly enhance biochemical sensor sensitivity. Its unique coupling phenomenon has been analyzed and utilized to boost the device's refractive index sensitivity to 967nm/RIU. Through optimization of the coupling strength between the two micro-resonators, further improvement in refractive index sensitivity up to 1100nm/RIU has been achieved. This mode coupling strategy allows us to design robust, thick-walled micro-tube sensors with ultra-high sensitivity which is useful in practical biochemical sensing applications

a) Electronic mail: taoling@umich.edu
b) Author to whom correspondence should be addressed. Electronic mail: guo@umich.edu.

Biological and chemical sensors play a significant role in a wide variety of applications related to everyday human life. Various sensing mechanisms based on the electrical[1], mechanical[2] and optical[3] properties of these devices have been explored to detect bio-chemical species. Optical-based sensors offer a number of attractive advantages, such as the capability of nondestructively detecting biochemical molecules, immunity to the electromagnetic interference, and the ability to be integrated with optical fibers for flexible remote detection. Among various optical sensing devices, optical micro-resonator based sensors.[4,5,6] have recently drawn a lot of attentions due to their unique advantage of reducing the device size by orders of magnitude without sacrificing the interaction length by virtue of the high quality (Q) factor resonances. With its compact size on the order of tens of microns the amount of sample needed for detection is significantly reduced. At the same time, the cavity-based resonant effect provides an increased effective interaction length and maintains sufficient sensitivity for detection. For example, optical microsphere cavities using whispering gallery mode (WGM) resonances have been demonstrated to respond to a monolayer of protein adsorption[7]. High Q thin-walled capillary cavities were capable of measuring a $2.8\times10^{-7}$ bulk refractive index change and a surface mass density change of 1.6 pg/mm$^2$.[8] Ultra-high Q ($>10^8$) micro-toroid cavities were able to detect single molecular binding events[9]. In these cavity-based sensing devices, the device performance is mainly determined by the device refractive index sensitivity and Q factor of the resonance mode. The device sensitivity is defined as the change of resonant wavelength with respect to the change of refractive index of the solution. The devices' Q factors are normally limited by the surface roughness which can be improved by optimization of the associated fabrication process. However, the device sensitivity is primarily determined by the spatial overlap between the optical field and liquid core region, which is a direct result of the cavity design. Therefore, it is of great importance to design a proper cavity that possesses a strong spatial overlap between optical field and liquid core regions in order to realize a highly sensitive bio-chemical sensor.

Several approaches have been proposed and realized to improve sensitivities of optical micro-resonators based sensors. O. Gaathon et. al have improved the device sensitivity by coating the surface of the resonator with a relatively high refractive index nano-layer[10]. With the help of high refractive index layer, the evanescent field of the coated microsphere extends further into the surrounding medium which contains the molecules to be detected. However, the evanescent field based sensors only have a limited sensitivity of up to 100nm/RIU. Previously, our group demonstrated a prism-coupled method to excite certain higher-order resonance modes with high refractive index sensitivity (around 600nm/RIU) in a silica micro-tube resonator[11]. Theoretical studies have shown that further increase of the order of the resonance modes will not necessarily yield further improvement of the device refractive index sensitivity[12]. M. Sumetsky et. al. reported using a very thin wall capillary tube to push more of the optical field into the liquid core region in order to realize much higher sensitivity. Sensitivity as high as 800nm/RIU in a bulk refractive index sensing experiment has been demonstrated [13]. However, the fragile, thinned wall hinders the applicability of such a device. G. Oh et al have theoretically proposed hybridizing a surface plasmon resonance (SPR) mirror structure with an InP-based triangular resonator. Due to the large phase shift in the SPR mirror, a significantly enhanced sensitivity of 930nm/RIU has been realized theoretically[14]. However hybridizing metal on a dielectric resonator could significantly increase the optical loss, which could lead to a poor detection limit.

In this work, we proposed using a liquid-filled micro-tube possessing a low-index inner coating to effectively behave as a coupled micro-resonator to greatly increase the refractive index sensitivity. This coupled resonator is formed by isolating a liquid cylinder resonator and a silica micro-tube resonator by an inner coated low refractive index material. Although these two micro-resonators are isolated by a low refractive index material layer, optical modes still can couple to each other and this coupling produces an anti-crossing behavior, which indicates strong mode coupling. Due to this strong coupling nature, under certain conditions, most of the optical field will be located in the liquid region and which results in an ultra-high bulk refractive index sensing sensitivity of 967nm/RIU. The sensitivity was further increased to 1100nm/RIU by optimizing the coupling strength between the two micro-resonators. These robust and sensitive resonator sensors have great potentials as transduction elements in practical bio-chemical sensors.

The proposed coupled micro-resonator is shown in the fig.1 (a) and it can be treated as a four-layer system (from inner to outer) in cylindrical coordinate: the first layer ($0<R<R_1$) is the liquid core with refractive index $n_1$, the second layer ($R_1<R<R_2$) is the coated low refractive index film layer with refractive index $n_2$, the third layer ($R_2<R<R_3$) is the silica tube layer with refractive index $n_3$, and the fourth layer ($R>R_3$) is the air with refractive index $n_4$. Such a resonator can be decomposed into two individual resonators: a silica micro-tube filled with a low index material which is shown in fig.1 (b) and a liquid micro-cylinder surrounded by a low refractive index medium which is shown in fig.1 (c). These two resonators have their own independent resonance modes. Under certain conditions, the independent resonance modes become hybridized to form a new coupled mode. This coupling phenomenon can be understood in terms of a 2x2 Hamiltonian matrix[15]:

$$H = \begin{pmatrix} E_1 & V \\ W & E_2 \end{pmatrix} \quad (1)$$

where $E_1$ and $E_2$ are the complex energies of the uncoupled system, and W, V are the coupling constants between two different states. The eigen-values can be expressed in the following form:

$$E_\pm = \frac{E_1 + E_2}{2} \pm \frac{1}{2}\sqrt{(E_1 - E_2)^2 + 4WV} \quad (2)$$

This coupled resonator system can also be studied in cylindrical coordinates. The time-independent field distribution for the resonance mode in the cavity can be separated into a radially-dependent component and an azimuthally-dependent phase term, R(r)exp(imφ), where R(r) is the amplitude of the axial magnetic (TE) or electrical (TM) modal field and m is the azimuthally quantization number. For simplicity, we only consider TM mode in the coupled-cavity system and therefore only three components: Ez, Hr and Hφ need to be considered. Hr and Hφ can both be expressed in terms of Ez. The radially-dependent field Ez can be expressed by Bessel functions in the following form [16]:

$$E_z(r) = \begin{cases} B_1 J_m(k_0 n_1 r) & 0 \leq r < R_1 \\ [B_2 J_m(k_0 n_2 r) + B_3 N_m(k_0 n_2 r)] & R_1 \leq r < R_2 \\ [B_4 J_m(k_0 n_3 r) + B_5 N_m(k_0 n_3 r)] & R_2 \leq r < R_3 \\ B_6 H_m^{(1)}(k_0 n_4 r) & R_3 \leq r < +\infty \end{cases} \quad (3)$$

where $J_m$, $N_m$ and $H_m^{(1)}$ are the $m$ th Bessel function, Neumann function and Hankel function of the first kind, respectively. By matching the boundary conditions at the liquid/polymer, polymer/silica and silica/air interfaces, we can obtain the eigen-equation, which can be used to calculate the resonance wavelength as well as the resonance Q factor. In all performed simulations,, the following parameters were used: inner radius $R_2$=134 μm, outer radius $R_3$=166 μm, the wall thickness is $d_1=R_3-R_2$=32μm, inner coating layer thickness $d_2=R_2-R_1$=1μm, liquid refractive index of $n_1$=1.32+0.0001i (water[17]), inner-coating material's refractive index of $n_2$=1.275(Teflon AF [18]), silica refractive index of $n_3$=1.45 and air refractive index of $n_4$=1.00 and azimuthal number m was fixed to be 700.

Fig.2 (a) shows a pair of resonance mode wavelengths in this coupled micro-resonator system as a function of detuning the refractive index of the liquid core. There are three significant features found in fig.2 (a): (i) With the liquid refractive index in the range of 1.32 to 1.331, the upper-branch mode resonance wavelength changes slowly, while the lower-branch mode resonance wavelength changes more rapidly, (ii) with the liquid refractive index in the range of 1.331 to 1.34, both upper and lower branches of the resonance wavelength show the opposite trends of the resonance wavelength changes with the liquid refractive index in the range of 1.32 to 1.331, and (iii) with the liquid refractive index at 1.331, there is a clear anti-crossing behavior where two resonance wavelengths come close but then repel each other. This anti-crossing phenomenon is well known to be an indicator of strong coupling and has been observed in many coupled systems, such as a quantum-dot coupled with micro-cavity [19, 20] and a plasmonic cavity coupled with a dielectric-cavity[21]. Additionally, resonance Q factors as the function of detuning of the refractive index of the liquid core has been shown in fig.2 (b). A crossing behavior in the resonance Q factor locates at the same liquid core refractive index position when the resonance wavelength showing the anti-crossing behavior. Away from the crossing position, one of the resonances has a higher Q factor ($\sim 10^5$) and the other resonance has a lower Q factor ($< 10^4$). This low Q factor resonance is mainly due to the water absorption loss at the 1550nm wavelength which will be further discussed in the later in the paper.

The device coupling phenomena [22] can also be investigated by the field intensity distribution in the cavity. Figure 3 shows the calculated field intensity distributions (normalized to the highest intensity of the field) in the cavity with different refractive indices of liquid: (I, II) correspond to the pair of resonance modes with the a liquid refractive index 1.324 (i.e. away from the anti-crossing position), (III, IV) correspond to the modes with a liquid refractive index of 1.331 (i.e. at the anti-crossing position) and (V, VI) correspond to the modes with a liquid refractive index of 1.338 (i.e. away from the anti-crossing position). Figure 3 (I, III, V) and (II, IV, VI) depict the field distribution of the resonance mode of lower and upper branches in fig.2 (a), respectively. It is clear that almost the entire the optical field in mode I (or VI) is confined in the liquid region (it can be called a liquid-like mode), and there is only a small part of the field in the silica tube region. Such a resonance mode shows great potentials for highly sensitive biochemical sensing. In contrast, for mode II (or V), most of the field is confined in the silica region (it can be called a solid-like mode), and only a small part is in the liquid region. Mode III and IV are the strongly hybridized modes and show a very similar field distribution in the cavity. When the liquid refractive index is detuned across the anti-crossing region, it can be clearly seen that the liquid mode and the solid mode shift from one branch to the other. This is clear evidence of strong coupling: the resonance mode exchanges its mode patterns as well as its energy while passing through the anti-crossing region [23]. Away from the anti-crossing region, optical modes are not strongly hybridized with each other, as shown in fig.3 (I, II) and (V, VI). They are either liquid-like modes or solid-like modes. The liquid-like modes (fig.3 I andV I) have more than 84% of the optical field located in the liquid region and show a great potential for bio-chemical sensing applications. Additionally, a part of the optical field extends into to the silica region (32um thick) and is still well-confined by total internal reflection at silica-air interface. Such a resonance mode shows that the field intensity exponentially decayed in the air region and can be effectively excited by tapered fibers or optical prisms [11] from outside of the micro-tube.

As mentioned above, the liquid-like modes could be excellent for bio-chemical sensing applications because of the optical field in these modes have a strong spatial overlap in the solution solution. The device biochemical sensing sensitivity (S) is defined as the change of the resonance wavelength with respect to the change of the liquid refractive index and it can be expressed as $S = \delta\lambda / \delta n_{liquid}$. The refractive index sensitivity of the device can be ascertained by taking the differential of the curves showing the resonance wavelength with respect to the refractive of the liquid in fig. 2 (a) .The sensitivities of the paired modes as a function of refractive index of liquid core are plotted in the fig.4. With the liquid refractive index in the range from 1.32 to 1.325, the upper-branch mode's sensitivity remains relatively low value (<100nm/RIU), while the lower-branch mode sensitivity can reach relatively high value (>800nm/RIU). With the liquid refractive index in the range from 1.325 to 1.338, the sensitivity of the upper-branch mode starts to gradually increase, at the same time the sensitivity of the lower-branch mode starts to gradually decrease. They start to show a crossing behavior at refractive index of 1.331 (the strongest coupling position), which means that these two modes have a similar portion of the optical field in the liquid region and this agrees with the plotted field patterns in the fig.3. After passing through the strongest coupling position, the upper-branch mode sensitivity continuous to grow and reaches relatively high values, meanwhile sensitivity of the lower-branch mode continuous to decrease to relatively low value (< 100nm/RIU). Due to the coupling nature, the highest device sensitivity is located at a specific refractive index range. In the fig.4, there are only two regions which have high refractive index sensitivity: one is located around n=1.325 and another one is located at n=1.338.

The highest refractive index sensitivity in the thick silica wall device can reach ~ 967nm/RIU. This number is higher than the results reported from a capillary tube with an ultra-thin wall[13]. This coupled micro-resonator method is an effective method to shift the optical field into the liquid region to increase the sensing sensitivity compared to single micro-resonator case[11,13], however, it limited the high device sensitivity to a particular refractive index range. This limitation could be compromised by future increasing the inner coating layer thickness which will widen the high sensitivity window. This will be discussed further later in the paper.

Further improvement of the device sensitivity can be realized by increasing the overlap between the optical field and the liquid solution. This overlap in our coupled micro-resonator system can be increased by reducing the coupling strength between these two individual resonators. One method to reduce the coupling strength is to decrease the refractive index of the material in the coupling gap. Fig.5 (a) shows the device refractive index sensitivity as the function of the inner-coating layer refractive index at a fixed layer thickness d=1μm. The sensitivity can greatly increase with the reduction of the refractive index of the inner coating layer, and it can reach > 1100nm/RIU when the inner coating layer refractive index is lower than 1.20. Another way to reduce the coupling strength is to increase the coupling gap distance between two resonators. In the proposed coupled micro-resonator system, this can be achieved by increasing the inner-coating layer thickness. Fig.5 (b) shows the sensitivity increased with the increase of the inner-coating layer thickness with a fixed inner-coating layer material refractive index (n=1.275). When the inner coating layer's thickness is ~ 2.5μm, the sensitivity can be as high as 1070nm/RIU. With the increase of the device's sensitivity, it also widens the window for high sensitivity refractive index sensing. We can clearly find from the inset figure in fig. 5(b) that the high sensitivity window becomes relatively flat and wide compared to the window in the fig. 4.

With this enhanced sensitivity, it can greatly help to improve the sensor detection limit (DL). The detection limit of the micro-resonator based sensing device is not only related to the sensitivity of the resonant mode, but also related to the Q factor. Suppose we can distinguish one twentieth of the resonant peak width, then the bulk refractive index detection limit (DL) can be expressed: $DL = \delta\lambda/(20S) = \lambda/(20SQ)$, where S is the device bulk refractive index sensing sensitivity, $\delta\lambda$ is the full-width at half maximum of the resonance peak, λ is the resonant wavelength and Q factor is equal to $\lambda/\delta\lambda$. Although the device bulk refractive index sensitivity can be as large as 1100nm/RIU, the device's DL is still ~ $7.3 \times 10^{-6}$ (refractive index unit) which is mainly limited by its Q factor (~ $10^4$). We believe that the water absorption loss is the dominant term in determining the overall Q factor of the resonance mode. Fortunately, water has very low absorption at visible or near visible wavelength. By switching the working wavelength from 1550nm wavelength range to near visible wavelength (e.g.800nm) range, water absorption loss can be reduced by two orders of magnitude. The Q can be as high as $4 \times 10^6$ at 800nm wavelength range with the bulk refractive index sensing sensitivity up to 600nm/RIU and the resulting detection limit would be ~ $1.3 \times 10^{-8}$ (refractive index unit). By combining the high Q factor and high sensitivity contributed by the mode coupling strategy, the coupled micro-resonator sensors could produce sensitivities of practical importance.

In conclusion, a coupled micro-resonator is proposed in a liquid-filled, low-refractive index material inner-coated silica micro-tube. Its unique coupling phenomenon has been analyzed and utilized to boost the device's index sensing sensitivity to 967nm/RIU in very thick wall silica (32μm) micro-tube. By tuning the coupling strength between two coupled resonators, the sensitivity as high as 1100nm/RIU has been theoretically predicted. This mode coupling method is a highly effective way to increase the overlap between optical field and liquid solution even in very thick wall micro-tube resonators. It provides an excellent solution for producing robust and highly sensitive biochemical sensors.

Figure Captions:

Fig.1 (a) A schematic of a low refractive index film inner-coated silica micro-tube sensor filled with liquid, (b) a silica micro-tube resonator filled with a low refractive index material and (c) a liquid cylinder resonator surrounded by a low refractive index material

Fig.2 (a) Anti-crossing behavior of the resonance wavelengths when the refractive index of liquid is ~1.331, (b) the Quality factor shows a crossing behavior when the refractive index of liquid is ~ 1.331
.

Fig.3 Calculated field intensity distributions in the coupled micro-resonator with the different refractive indices of liquid, (I, II) with the liquid refractive index of 1.324, (III, IV) with the liquid refractive index of 1.331, (V, VI) with the liquid refractive index of 1.338

Fig.4 Sensitivity of two hybridized resonance mode as a function of liquid refractive index

Fig.5 (a) Sensitivity as a function of the inner-coating layer material refractive index (with fixed thickness d=1μm). (b) Sensitivity as a function of the inner-coating layer material thickness (with fixed refractive index at n=1.275) and the inset figure shows the sensitivity of the two hybridized modes as a function of liquid refractive index at inner-coating layer thickness 2μm

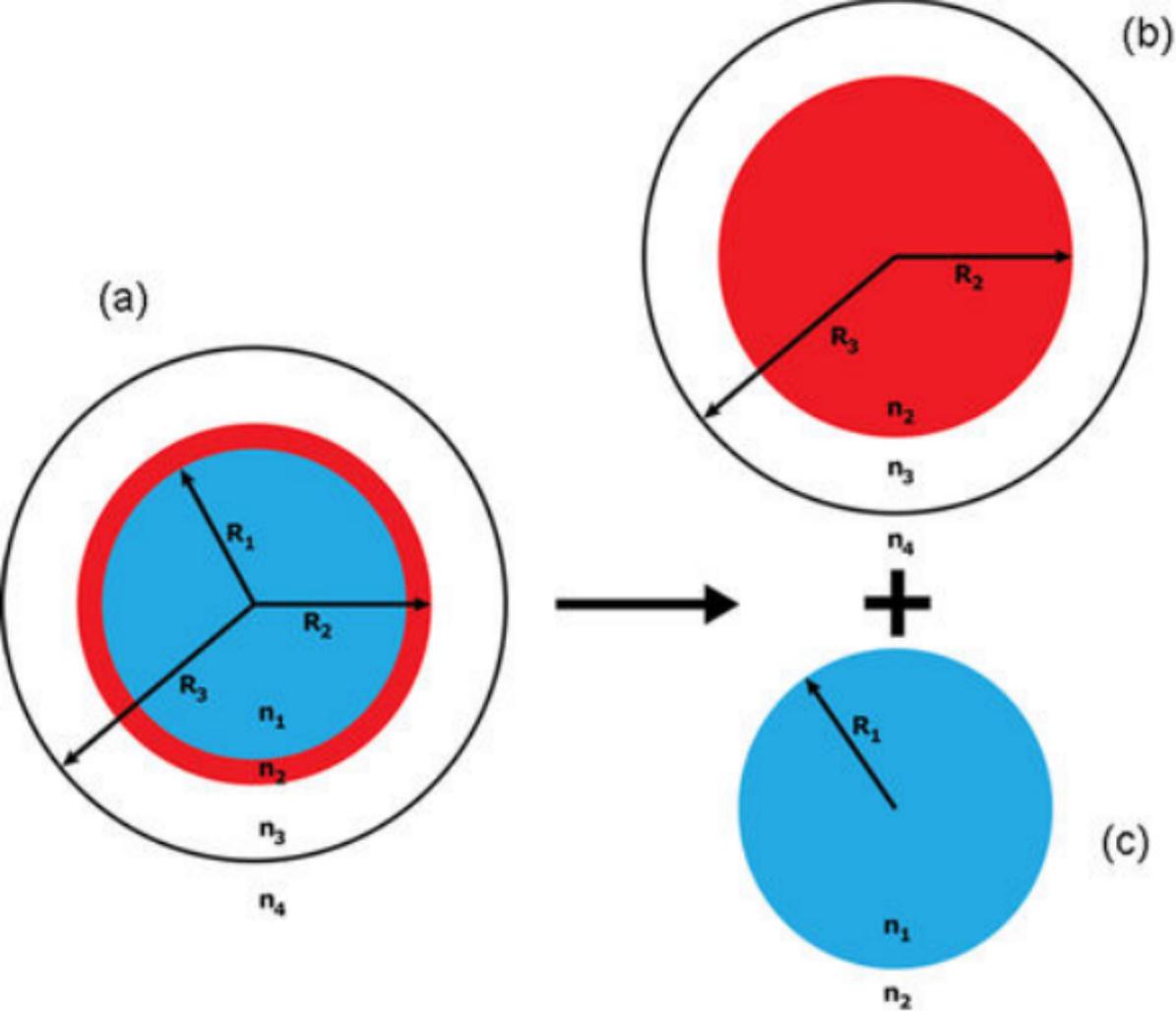

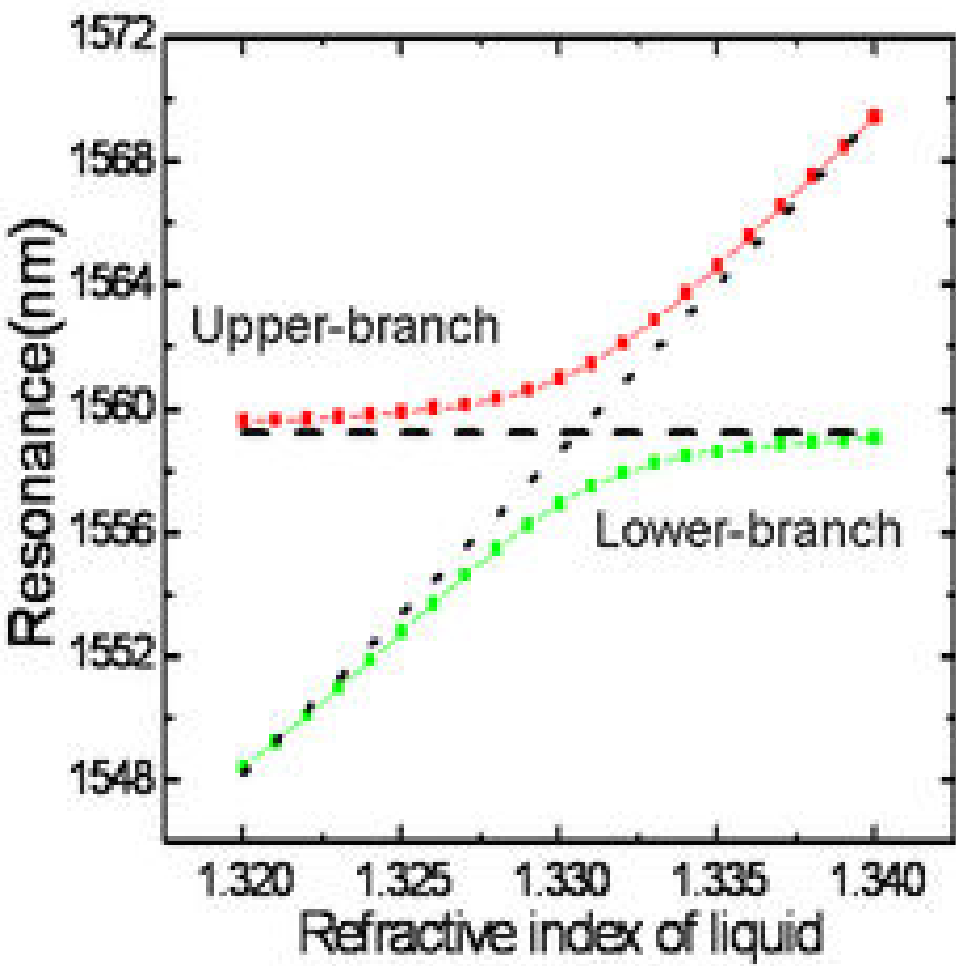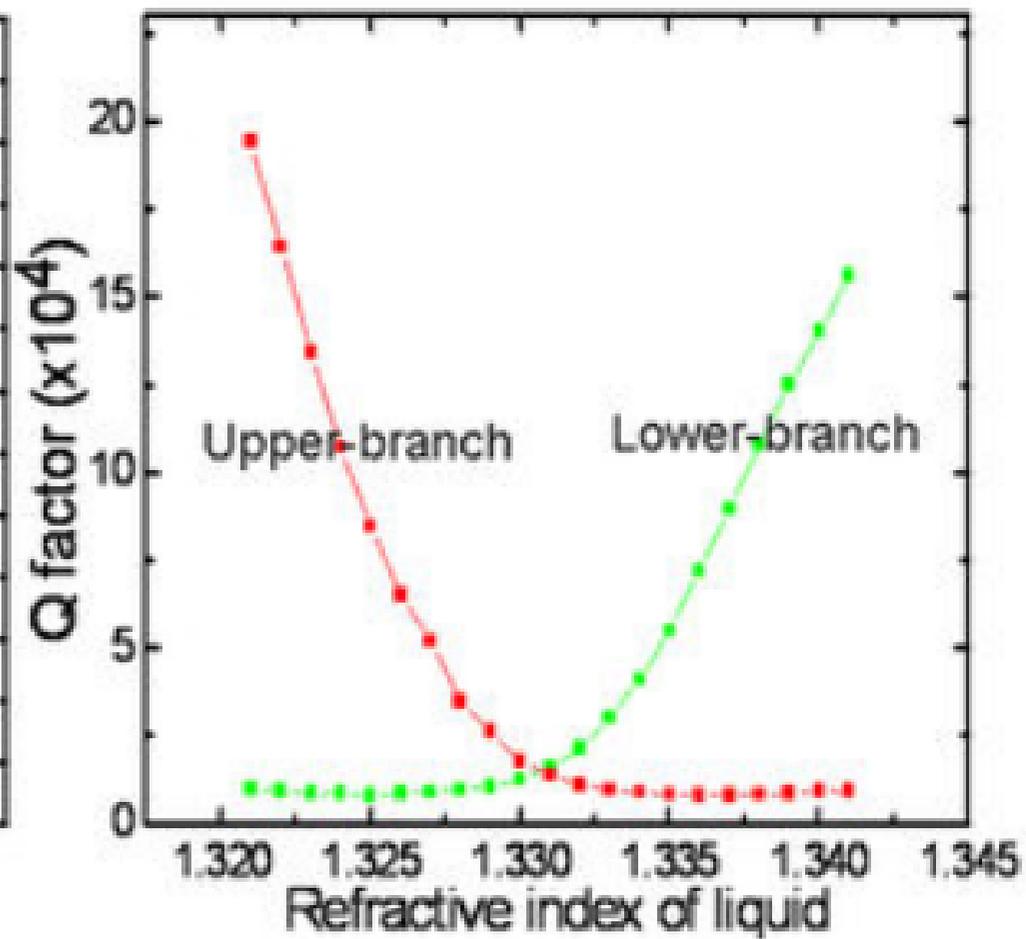

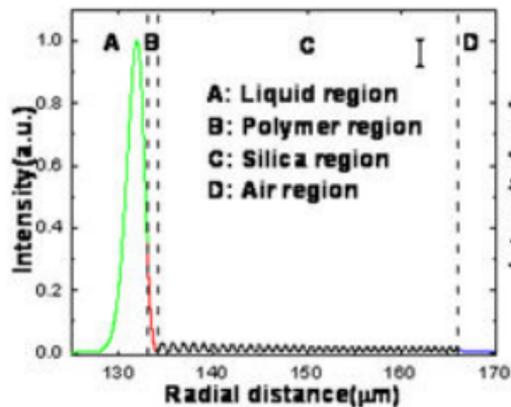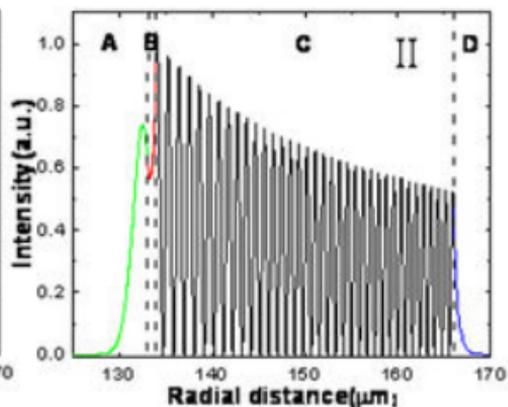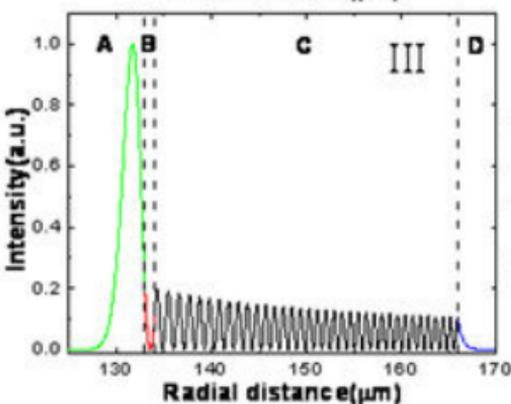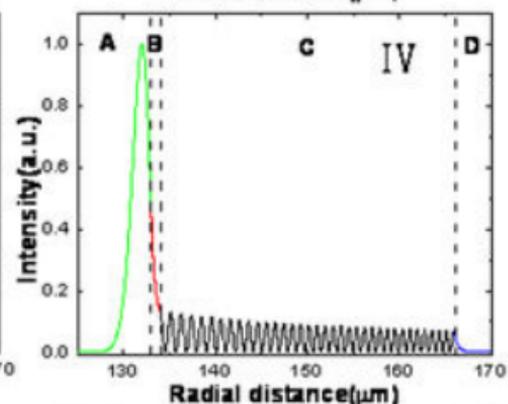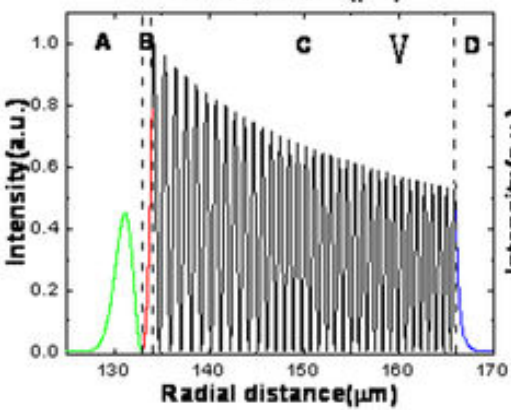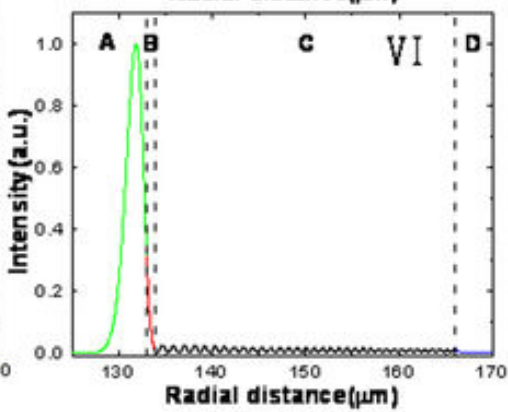

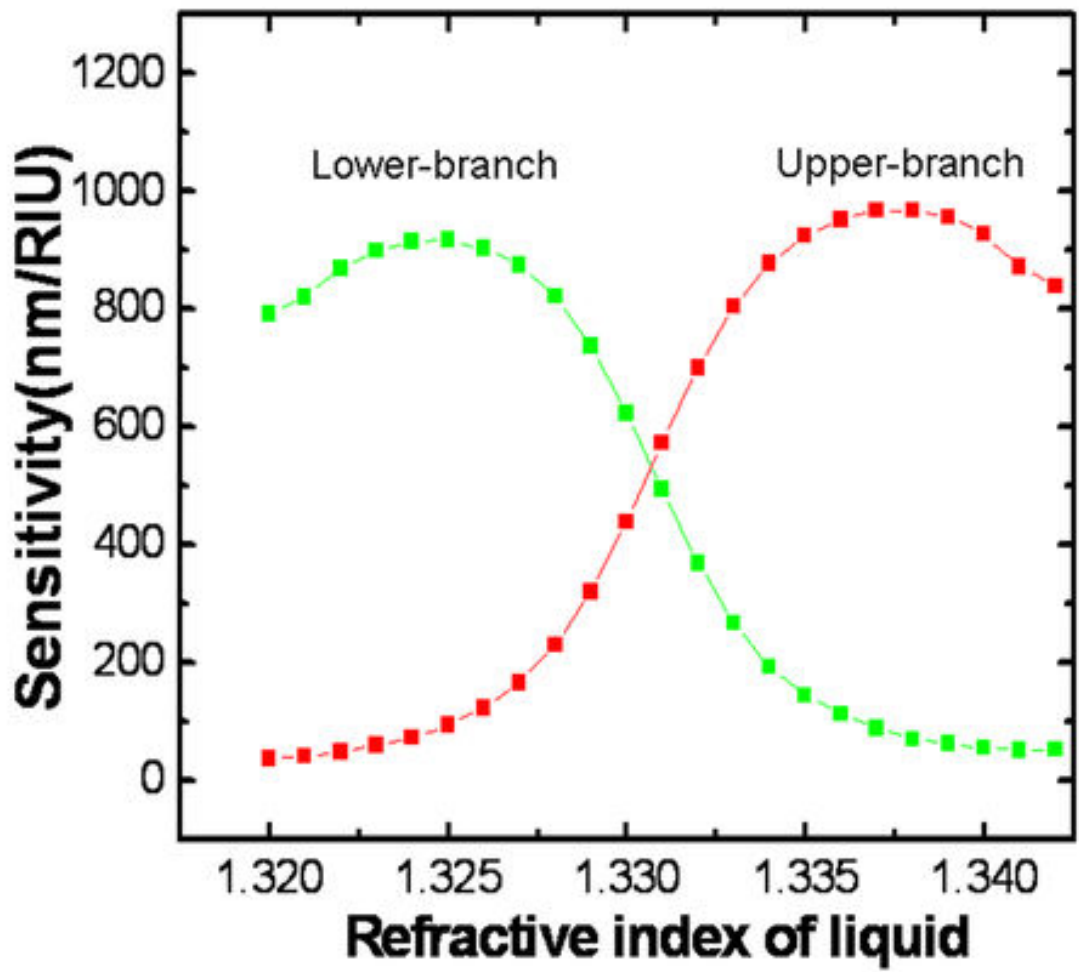

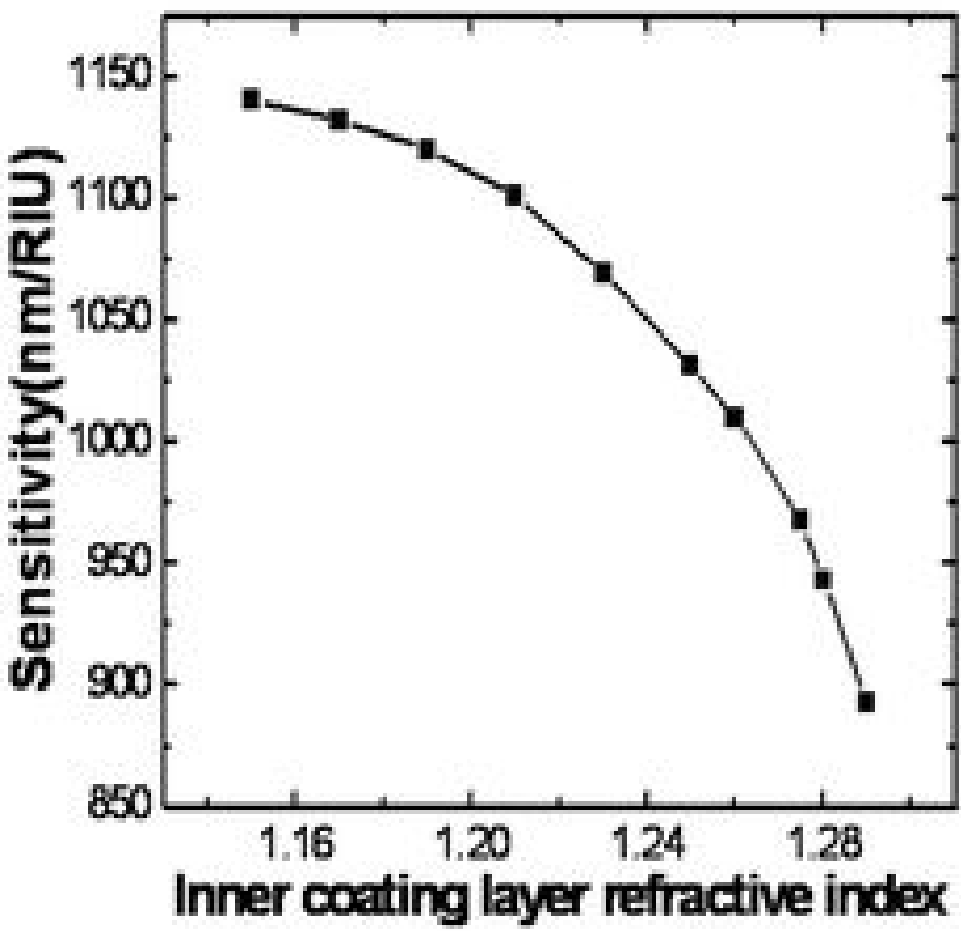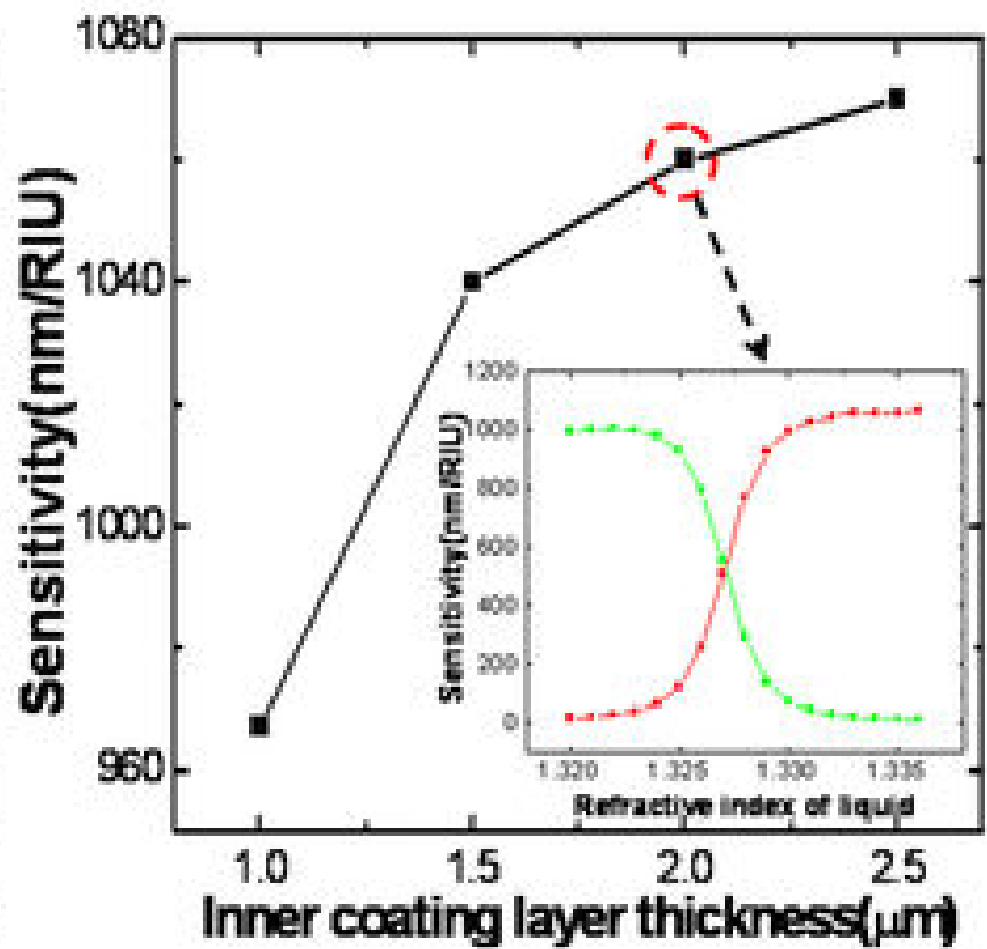